**SETI surveys of the nearby and distant universe employing wide-field radio interferometry techniques**

M.A. Garrett[a,b*]

[a] Jodrell Bank Centre for Astrophysics (JBCA), School of Physics & Astronomy, The University of Manchester, Oxford Road, M13 9PL, UK

[a] Leiden Observatory, Leiden University, P.O. Box 9513, NL-2300 RA, Leiden, The Netherlands.

* Corresponding Author

**Abstract**

Search for Extra-terrestrial Intelligence (SETI) research has recently undergone a major rejuvenation with initiatives like the Breakthrough Listen project conducting new systematic surveys of the nearest stars and galaxies. Current SETI surveys typically focus on the analysis of raw voltage data obtained from either large single dishes (e.g. GBT, Parkes etc.) or beam-formed (typically short-baseline) arrays (e.g. ATA, and in the future MeerKAT). This permits a standard and traditional SETI analysis of the data to be made with good time and frequency resolution. Radio interferometers, separated by distances > 10 km can also provide some interesting opportunities for future SETI searches. For some time, it has been recognised that Very Long Baseline Interferometer (VLBI) arrays are less affected by radio frequency interference (RFI), greatly reducing the number of false positives that arise from terrestrial and satellite transmissions. In this paper we highlight other advantages including the presence of multiple interferometer baselines in an array - these provide an important level of redundancy and additional confidence (verification) of faint and potentially transient signals. The need for high time and frequency resolution in SETI is also well matched to wide-field VLBI observations employing the latest VLBI software correlators. These can generate high time and frequency resolution data that permits an analysis of the full field-of-view (limited only by the primary beam response of the individual antenna elements). Using various wide-field and spectral line VLBI analysis techniques, thousands of potential SETI targets can therefore be studied simultaneously. These range from nearby galactic stars (identified via their Gaia proper motions) and distant extragalactic systems.

By making use of archive EVN data, we explore how one might prepare and analyse VLBI observations from a SETI perspective. The data follow a traditional calibration process in which the response of the central bright calibrator (J1025+1253) is subtracted from the uv-data. The data are then phase rotated to two targets within the field of view - Gaia DR2 3883720981953003776 a galactic star and [RGD2013] J102550.22+125252.82 a galaxy with a measured redshift of z=0.14. Searching for a SETI signal in the image plane has the important advantage that the signal location on the sky is likely to be invariant – this is a useful constraint when potentially almost everything else could be changing (e.g. frequency drift due to Doppler accelerations, temporal variability etc). In addition, the position invariance is an excellent discriminator against false-positives (terrestrial RFI) By analysing the statistics of images generated for all available frequency channels, we place coarse upper limits on any SETI signals from the two SETI targets, and note that while the e.g. EIRP (Equivalent Isotropic Radio Power) associated with the limits we have placed on the galaxy are comparable to the energy resources of a Kardashev Type II civilisation (~ $10^{26}$ W), a distributed array of coherent transmitters with excellent forward gain, could reduce this to more modest levels. We therefore argue that targeted observations of extragalactic sources are also merited by SETI (interferometer) surveys. We caution that this is very much a preliminary analysis, and that the complete range of analysis strategies using interferometers still needs to be fully developed. Looking towards the future, we content that in the event that a SETI candidate *is* discovered, radio interferometers distributed on scales of 1000's of km will also be able to pin-point the location of the extra-terrestrial transmitters with (sub-)milliarcsecond precision – this may be crucial in understanding the characteristics of the platform on which the source is fixed, and potentially the nature of the civilisation or other entity responsible for generating the signal. If SETI signals are moving in space, VLBI techniques can detect objects at a distance of 1 kpc with velocities > 0.01c via repeated observations separated by only one day. The motion of vehicles with velocities similar to the Voyager spacecraft can be detected within 1 year.

**1. Introduction**

The Search for Extra-terrestrial Intelligence (SETI) is a field of research that is about to celebrate its 60[th] anniversary as a scientific pursuit [1, 2]. So far, no SETI signals have been detected, although a new initiative, Breakthrough Listen, has embarked on a major new campaign to systematically survey the nearest stars and galaxies [3]. Current SETI surveys typically focus on the analysis of raw voltage data obtained from either single dishes or beam-formed (typically short-baseline) arrays. While beam-formed instruments can target specific sources in the field, it is not always fully appreciated that single dish observations do not.

The first SETI VLBI observations using the Australian Long Baseline Array [4] demonstrated the advantages of an interferometer in mitigating the effects of human-made radio frequency interference (RFI). In this paper, I present the other advantages that an interferometry-based analysis poses for future SETI surveys, especially for interferometer arrays distributed on scales of 10-10000 km (VLBI). In particular, the high frequency and time resolution required for SETI searches, also permits a wide-field VLBI approach to be employed. In Section 2 I summarise the main advantages of the interferometric approach presented here to SETI. In Section 3, I describe the analysis of an archive EVN (European VLBI Network) observation of a field that includes a bright extra-galactic radio source, demonstrating how we can search for SETI signals from many other sources in the field, including galactic stars, and both nearby and distant extragalactic systems. In Section 4, we arrive at some conclusions and suggestions for future work is proposed in section 5.

## 2. Advantages of an Interferometric approach

Radio interferometers offer several advantages over single-dishes and beam-formed arrays for SETI research. We present some of the most important factors in this section.

*2.1 Sensitivity and Field-of-view*

Radio interferometers such as VLBI arrays often combine together the sensitivity of the largest radio telescopes in the world. Potentially, they can achieve the highest sensitivity of any radio telescope on the planet. Historically, this has been achieved at the expense of field-of-view since for a given data integration time and data frequency resolution, the field-of-view scales inversely with the square of the maximum baseline length. Today, modern software correlators [5] can generate enough time and frequency resolution that very sensitive observations can detect multiple sources across the primary beam of individual VLBI antennas [6]. As we shall see in section 3, SETI observations using VLBI data can therefore extract and target the many thousands of galactic and extragalactic sources simultaneously present in the same field of view.

*2.2 Detection significance, confidence and redundancy*

A major issue with the detection of SETI signals is likely to be the significance of the detection and confidence in the result. This is particularly true for faint sporadic or non-repeating signals. A radio interferometer of N antennas, generates $N(N-1)/2$ *independent* baselines. Since SETI signals are very likely to be unresolved, a signal detected in one baseline must also be detected in another. For an array of 10 antennas, there are therefore a total of 45 independent baselines available to verify any given event. The history of SETI is littered with claims and counter-claims of possible signal detections (with perhaps the "wow!" signal being the best well-known). The redundancy and greater confidence made available via interferometry is an important advantage over single-dish and beam-formed instruments. In addition, interferometers make images, and one of the few invariants in SETI signal searches is that the location on the sky is very likely fixed (see also section 6). This is not true of other potential SETI signal characteristics which include temporal variability and frequency changes due to the expected Doppler drift. The position of a SETI signal is invariant, and this can be a very useful constraint also in the data analysis in weeding out residual false-positives. The concept is shown in Fig 1. below.

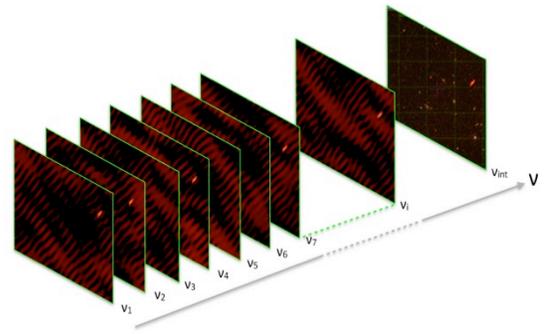

Fig. 1. The position of the SETI signal is invariant in its location on the sky. While everything else might be changing (e.g. central frequency or temporal variability) the signal location remains fixed, at least within the duration of a short observation.

*2.3 Radio Frequency Interference (RFI)*

The advantages of radio interferometers in mitigating against human-made RFI is well documented e.g. [4]. Basically, long-baseline radio interferometers are significantly less affected by spurious RFI than either single dish telescopes or short-baseline beam-formed arrays. This is particularly the case for long baseline arrays distributed on scales

> 30km. In this case, sources of terrestrial RFI are unlikely to be common for such widely separated antennas. As a result, sources of RFI typically do not correlate between antennas. Even if the RFI sources are correlated (e.g. satellite transmitters), they will usually be located far from the pointing (phase) centre of the array. Such sources suffer from very high residual fringe (phase) rates (which scale with the baseline length), such that time-average smearing will lead to severe de-correlation of any common narrow-band RFI signal [7].

## 3. SETI search demonstration using archive VLBI data

I present a demonstration of how one might generally conduct a search for SETI signals, using VLBI. The demonstration data were extracted from the EVN archive (see http://jive.eu/archive-introduction). The approach is different from previous analyses [4], recognising that while the properties of a SETI signal are likely to vary in terms of time and frequency (e.g. Doppler accelerations for narrow band signals), the single invariant that can more likely be relied upon is the SETI signal's fixed position on the sky (see Fig. 1). Our approach therefore involves the generation of large 3-D image cubes (each frequency and in principle time) in which the signal position on the sky is (presumably) invariant.

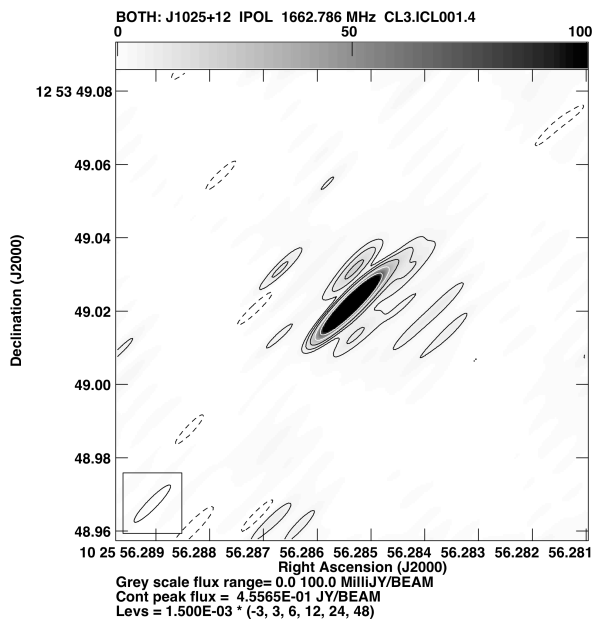

Fig. 2. Contour map and greyscale image of the calibrator J1025+1253. Contours are plotted at -3, 3, 6, 12, 24, 48 times the rms noise level of 1.5 mJy/beam.

### 3.1 Observations and data analysis

The EVN archive data set employed for this SETI demonstration is ED038 (PI Deller). The data were correlated at the Joint Institute for VLBI ERIC (JIVE), and processed by the automatic EVN pipeline. ED038 was chosen because the visibility integration time was 0.25 seconds with a frequency channel resolution of 31.25 kHz over a total bandwidth of 128 MHz (LCP & RCP) – 1.595-1.722 GHz. Table 1 shows the main characteristics of the observational set-up.

Table 1. Observational set up.

| Parameter | Value |
|---|---|
| Frequency range | 1.59-1.72 GHz |
| Total bandwidth | 128 MHz (LCP & RCP) |
| Number of channels | 8192 (LCP & RCP) |
| Channel width | 31.25 |
| Time on source | 507 sec |
| Integration time | 0.25 sec |

The data included nine very short scans on the bright calibrator, J1025+1253, representing a total on source time of 507 seconds. The scans are irregularly spread over a period of exactly 2 hours. The source was observed on the 10$^{th}$ of June 2012. The data were only coarsely edited e.g. channels at the start and end of each of the 8 IFs, and periods when antennas were known to be down.

### 3.2 SETI VLBI analysis concept

A self-calibrated map of the J1025+1253 is shown in Fig. 2. The self-calibration process yielded small amplitude and phase corrections that were applied to the un-averaged multi-source data set. The first stage of the analysis specific to SETI was to subtract the response of the bright calibrator from the final corrected un-averaged data set using AIPS task UVSUB.

Thereafter, it was possible to generate 3-D image cubes (in frequency) for any location within the field of view (see Fig.3). This was performed by using the AIPS task IMAGR[1] to phase shift the data to the target of interest, and then to generate dirty images at that location for each channel, and for many short time intervals of duration 10 seconds. Fig. 3 shows the concept of generating frequency (and time) cubes for one of the targets (see section 4).

---

[1] Note SETI VLBI observations made today can execute the phase shifting step earlier in the process, during the initial software correlation.

Because of limitations in disk resources, and since at this stage we are still in the proof-of-concept phase, we generated each frequency channel image using the full 510 seconds of on-source data. Similarly, each image generated over a period of a few seconds, averages across the full bandwidth (128 MHz). This limits the types of signal we can detect (for further discussion see section 4). The image cubes were converted into continuous movies so that the data could be first visually inspected.

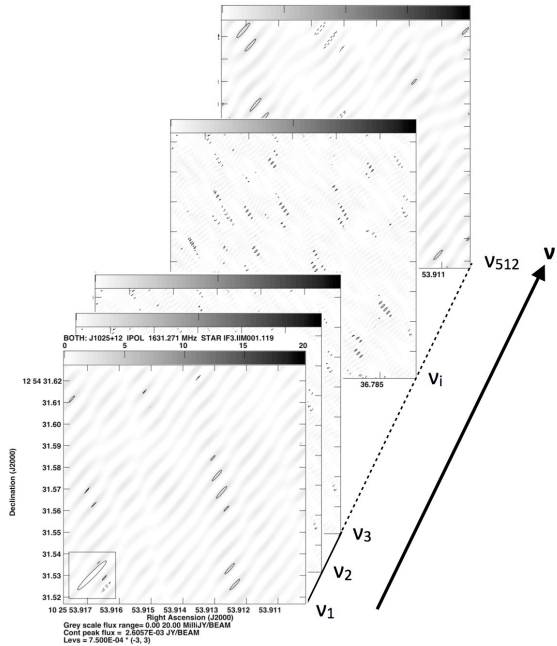

Fig. 3. Contour maps and greyscale images of a region centred on the galactic star, Gaia DR2 3883720981953003776, one of the SETI targets in the field of J1025+1253. Images are produced for each of the 31.25 kHz frequency channels (up to 512 channels, $\upsilon_1 - \upsilon_{512}$). Each channel images uses all the data (multiple scans with a total observing time of 502 seconds).

## 4. Results

Clearly there are many potential SETI targets in the field centred on J1025+1253 (see Fig. 4). This preliminary analysis was restricted to sources located within 1-2 arcminutes of J1025+1253 due to concerns about time and bandwidth smearing (a more modern data set generated by a software correlator would have access to a field-of-view only limited by the primary beam of the individual antennas). We chose two possible "SETI targets" within this limited field-of-view (see Fig. 2 above) - Gaia DR2 3883720981953003776 [8] a galactic star and [RGD2013] J102550.22+125252.82 a galaxy with a measured redshift of z=0.14107 [9]. We note in passing, that targeting stars with large proper motions is aided by the availability of Gaia measurements.

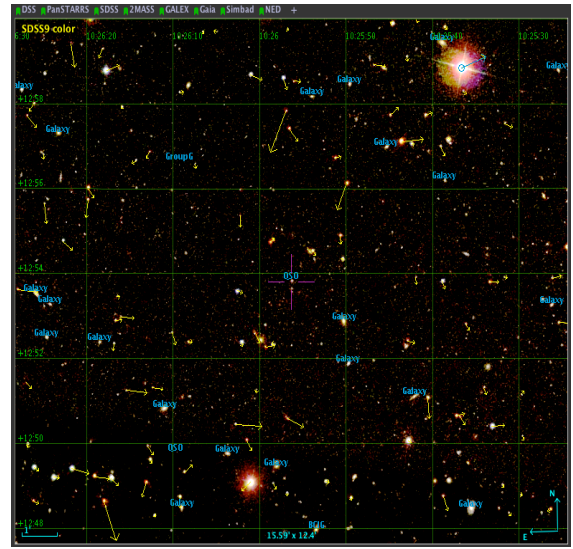

Fig.4. The SDSS optical field associated with J1025+1253 (see red cross "QSO" label in the centre of the field). Sources with yellow arrows are galactic stars with measured proper motions via Gaia (DR2) [8]. Other extended objects are identified as relatively nearby galaxies. More distant extragalactic objects are faint and extended on scales less than a few arcseconds. All these sources are potential SETI targets, and all are accessible via a wide-field interferometric approach.

In this preliminary paper we confine ourselves to the more sensitive frequency cubes but note that it is also possible to generate the same cubes but as a function of time. For DR2 3883720981953003776 a search through the frequency cubes show no signals greater than 0.009 mJy (4-sigma). With a measured parallax of 0.9769+/-0.1457 the star is located at a distance of ~ 1024 pc from Earth. At this distance, the flux density limits in the frequency and time cubes correspond to an Equivalent Isotropic Radio Power (EIRP) of $3.5 \times 10^{16}$ W.

At the galaxy position similar flux density limits are measured, corresponding to EIRP upper limits of $1.4 \times 10^{28}$ W. These EIRP power levels are at a level associated with a Kardashev Type II scale civilisation ($\sim 10^{26}$ W). Naturally these EIRP powers are large but directive antennas can reduce the power requirements substantially e.g. Arecibo has a forward gain of 60dBI, and one can imagine a distributed array of transmitting antennas with a forward gain orders of magnitude larger than that.

Note that the fairly coarse frequency resolution of these observations (31.25 kHz) means our sensitivity to narrow-band signals is poor - a 3 kHz signal is diluted by a factor greater than 10 in our data. Being only sensitive to fairly broad signals (of order the channel width and greater), we can, however, neglect any effects due to possible Doppler accelerations between the target and observer. Finally, we note that these measurements are preliminary, and a more in-depth analysis would certainly provide more stringent upper limits.

## 6. Conclusions

SETI surveys employing long baseline interferometry is a technique worth pursuing further. The invariant nature of the position of the signal (irrespective of other varying signal characteristics) could be a useful discriminant against spurious false-positives which are already greatly reduced via the use of interferometry. It's possible that a combination of beam-forming and interferometric techniques could be optimum for SETI research conducted on short-baselines arrays, such as MeerKAT.

The next step for SETI VLBI is to make some test observations using software correlators with much higher spectral and time resolution. The spectral resolution of the data presented here greatly limits the EIRP constraints on narrow-band signals. This coupled with much longer integration times can make VLBI a very competitive technique for SETI research. The only obvious disadvantage of the SETI VLBI approach is the amount of data generated – basically that scales as $N(N-1)/2$ for a random array of N telescopes, compared to the beam-forming approach. For sparse arrays like VLBI, this disadvantage is probably offset but the other advantages. Future VLBI SETI surveys should probably target as many sources as possible in the natural field-of-view, including exotica such as extragalactic systems – even very distant systems.

We note that in the event that a SETI signal is discovered, VLBI will play an important role in characterising the nature of the transmitter and the platform it sits on. Figure 5 highlights the matter. With a resolution better than 1 mas, the orbital motion of a transmitter located on a planet with an orbital radius of 0.5 AU from its parent star can be easily detected. In addition, motion of a moving transmitter, perhaps located on a robotic spacecraft, traveling at 0.01c can be detected by repeated observations separated by 1 day. Motion of a much slower spacecraft (similar to the Voyager and New Messenger spacecraft) can be detected via repeated observations separated by about 1 yr.

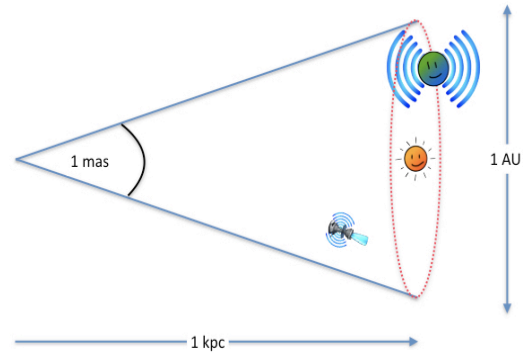

Fig.5. VLBI provides mas or sub-mas scale resolution – this can resolve the orbital motion of a SETI signal (orbiting around a star) at a distance of 1 kpc. Motion of an object travelling at 0.01c could be detected within 1 day.

**Acknowledgements**

MAG acknowledges support from an IBM Faculty Award.